# Structure of the Magneto-Exciton and Optical Properties in Fractional Quantum Hall Systems

Jun Zang* and Joseph L Birman

*Department of Physics, City College of New York, New York, N.Y. 10031*

## Abstract

We report calculated dependence of magneto-exciton energy spectrum upon electron-hole separation $d$ in Fractional Quantum Hall systems. We calculated the dependence of photoluminescence upon $d$, and we obtained the doublet structure observed recently. The Raman scattering spectrum around resonance is calculated: a robust resonance peak at $\nu = 1/3$ around gap value is reported.

Strongly correlated two-dimensional (2D) electron systems in a strong magnetic field reveal novel transport properties known as the Fractional Quantum Hall effect (FQHE) [1,2]. Recent optical experiments [3,4] discovered new, interesting properties which may provide useful information about the excitation spectra of the *incompressible* states [5] in FQHE systems. The PL experiments [3] showed very interesting energy anomalies, and a noticeable correlation between magnetic field dependent photoluminescence (PL) intensity modulation and the well known field-dependent transport measurements in $\sigma_{xy}$ and $\sigma_{xx}$. The behavior of PL intensity modulation was first interpreted by Wang, Birman, and Su [6] using a simple model. Other properties have been investigated theoretically [7–9]. However, the energy anomalies (doublets or triplets) and their temperature and magnetic field dependence are still not well understood. Recently Pinczuk *et al.* [4] observed the magneto-roton resonance peak around the gap value $\Delta$ in a Raman scattering (RS) experiment at $\nu = 1/3$ filling fraction. Usually, the RS efficiency is modeled so as to be proportional to the dynamic structure factor $S(Q,\Omega)$ of the conduction band electrons [10]. It is well known [11] that the oscillator strength of the magneto-roton mode is proportional to $Q^4$ , *i.e.* $S(Q,\Omega) \sim Q^4$. So the magneto-roton cannot be measured in a usual RS experiment for $Q = 0$. The only reason it can be measured is because of the resonance enhancement and appearance of the valence hole [10] in the intermediate state. So the magnitude of the RS peak from the magneto-roton mode is very sensitive to the incident photon energy $\omega_1$, as found in the experiment [4].

Since in the optical measurements, the valence hole will be strongly correlated with the electrons, the behavior of a many-body magneto-exciton [12,13] is crucial in understanding the optical properties of FQHE systems [8]. In this letter, we will report our numerical results obtained by exact diagonalization method for the properties of the many-body magneto-exciton, and the related optical properties of PL and RS in the FQHE systems. (i) We find that the magneto-exciton states can be classified into two "phases", which depend on electron-hole separation $d$. (ii) The PL spectrum dependence on $d$ is studied and the doublet structure in the PL spectrum is explained. (iii) the RS scattering spectrum in FQHE systems is first calculated and a resonance peak (at energy slightly above the gap value $\Delta$) is found.

The details of this calculation will be reported elsewhere [14]

Here we will simplify the problem by assuming the valence hole is also in the lowest Landau level. We will use the spherical geometry [15,16], so the electron-hole system is modeled as two "concentric" spheres with the same radius $R = \sqrt{S}$, where the total flux crossing the sphere is $2S$. In the lowest Landau level, there are $2S+1$ degenerate states [16] $|m\rangle$ with $m = -S, -S+1, \cdots, S$. The filling fraction is defined by $\nu = (N-1)/(2S)$. The electron-hole interaction can be written as $-\frac{1}{\sqrt{R^2|\vec{\Omega}_1 - \vec{\Omega}_2|^2 + d^2}}$, where $\vec{\Omega}_i$ is a unit vector in the radial direction denoting the electron (hole) position in the electron (hole) sphere. Here we use the magnetic length $l_0$ as the unit of length and $e^2/\varepsilon l_0$ as the unit of energy.

Fig.[1] shows the spectrum of the magneto-exciton built of 6 electrons (6e) and 1 hole (1h) with $2S = 12$, at different electron-hole distances $d$. We have also calculated the spectrum for 5e and 1h with $2S = 9$, and the same properties were found [14]. The spectrum of the magneto-exciton shows an interesting dependence on the distance between electron and hole layers, and can be classified into two regions depending on the distance $d$. In the region $d > d_c$, where $d_c \sim 0.8R$, the spectrum of the magneto-exciton has low energy branches which have parabolic-like dispersion relations. The origin of these magneto-exciton states can be understood most easily in the large $d$ limit. At $d = \infty$, the electron and hole are decoupled. The eigenstates are the direct product of many-body electron states and hole states. If we fix the total $L_z = M$, the Hilbert space is $\mathcal{H} \equiv \sum_{m=-S}^{S} \varphi_m \mathcal{H}^e_{\{M-m\}}$, where $\mathcal{H}^e_{\{M-m\}}$ is the subspace of the electron system with angular momentum $L^e_z = M - m$, and $\varphi_m$ is the wave function of the hole with angular momentum $l^h_z = m$. Without losing generality, we can set $M = 0$. For each eigenstate of $N+1$ conduction band electrons with $L^e = l$, there are $2l+1$ multiplet states $L^e_z = -l, \cdots, l-1, l$, and so the degeneracy of the many-body magneto-exciton states is $min(2l+1, 2S+1)$. Each many-body eigenstate of conduction electrons corresponds to a degenerate band for the many-body magneto-exciton at $d = \infty$. Turning on the electron-hole interaction adiabatically by decreasing $d$, the magneto-exciton obtains a finite renormalized mass due to e-h interaction. In the sense of the mapping: $L \sim k$(momentum), the magneto-exciton has approximately a parabolic

dispersion relation. This structure of the spectrum for low energy states survives with the increase of e-h interaction, until $d \sim d_c$. After $d < d_c$, the magneto-exciton goes into the strongly correlated region of electron and hole. In this region, the effect of e-h interaction is comparable to that of the e-e interaction. The many-body magneto-exciton in $d < d_c$ region can be characterized as the magneto-exciton in the "symmetric model" at $d = 0$. In this model [7,12], one set of eigen-states of magneto-exciton $|\bar{\Psi}_i^{ex}\rangle$ can be obtained directly from the eigen-states $|\Psi_i^e\rangle$ by $|\bar{\Psi}_i^{ex}\rangle = \sum_{m=-S}^{S} e_m^\dagger h_m^\dagger |\Psi_i^e\rangle$, where $e_m^\dagger$ and $h_m^\dagger$ are creation operators of (conduction) electron and (valence) hole respectively. This relation is shown in Fig. 1(a).

We can now use these results to calculate the optical response. In PL, the initial state is the magneto-exciton state which has $N+1$ electrons in the conduction band and 1 hole in the valence band; after recombination of an electron with the hole, the final state is the $N$ electron FQHE state. In the transition processes, the coupling between electron (hole) and photon is the $\vec{j} \cdot \vec{A}$ term. Decompose the vector potential as $\vec{A} = A_+\hat{e}_+ + A_-\hat{e}_- + A_0\hat{e}_0$, where $\hat{e}_+ = \hat{e}_1 - i\hat{e}_2$, $\hat{e}_- = \hat{e}_1 + i\hat{e}_2$, and $\hat{e}_0 \equiv \hat{e}_3$ are polarization vectors of the electromagnetic field. The electron-photon coupling term in the dipole approximation can be written as

$$H_{int} = c_1 A_+ \hat{\mathcal{L}}_{+1} + c_{-1} A_- \hat{\mathcal{L}}_{-1} + c_0 A_0 \hat{\mathcal{L}}_0 + h.c. \tag{1}$$

where $c_i$ are constants and $\hat{\mathcal{L}}_{i=0,\pm1} = \sum_m \hat{e}_m \hat{h}_{m+i}$. Previous studies [7,9] used only $\hat{\mathcal{L}}_0$ as the transition operator. However, as $\hat{\mathcal{L}}_0$ mainly couples with $\vec{A}_3$, the electromagnetic fields with this polarization will propagate along the $x-y$ plane and cannot be measured in the experiments. So the relevant transition operators are $\hat{\mathcal{L}}_{+1}$ and $\hat{\mathcal{L}}_{-1}$ [17]. However, we will also study the PL spectra due to $\hat{\mathcal{L}}_0$ just for comparison.

From the Fermi Golden rule, the PL spectra are given by

$$P_i(\omega) = M \sum_{m,n} e^{-\beta E_n^{ex}} \left| \langle \Psi_m^e | \hat{\mathcal{L}}_i | \Psi_n^{ex} \rangle \right|^2 \delta(\hbar\omega - E_n^{ex} + E_m^e) \tag{2}$$

For simplicity, we have assumed that only circularly polarized light is measured. Since $P_{-1}(\omega)$ is identical to $P_{+1}(\omega)$ we will let $P_1(\omega)$ in this paper represent $P_{\pm1}(\omega)$.

The spectra $P_{0,1}(\omega)$ has complex structure and dependence on the e-h interaction (dependence on $d$), originating from the dependence of magneto-exciton states on $d$ discussed above. Before discussion of the general features of these $d$ dependence PL spectra, let's first recall that in the limit of $d = 0$, the PL spectra due to $\hat{\mathcal{L}}_0$ has a single peak [7,12] $P_0(\omega) = \delta(\hbar\omega - E_h)$ which comes from the transitions between the corresponding states shown in Fig. 1(a), here $E_h$ is the "activation" energy of a single hole. And this certainly is not true for the $P_1(\omega)$, since $\hat{\mathcal{L}}_{\pm 1}$ does not commute with the Hamiltonian $H_{tot}$. However our numerical calculations shows that there is a dominant single peak for $P_1(\omega)$ at $d = 0$ for $T \sim 0.5k - 2K$, *i.e.* $T \sim 0.003 - 0.01(e^2/\varepsilon l_0)$. Other peaks are almost invisible although they are not zero.

At low temperature $T \sim 0.003 - 0.01(e^2/\varepsilon l_0)$, only low energy initial states are occupied, thus are relevant for the PL. From Fig. 1, we can see two kinds of initial states are important here: (i) lowest $L \sim 0$ states; (ii) the lowest branch states at $d > d_c$ and their correspondence at $d < d_c$. Because of angular-momentum (or momentum) conservation, the first kind of states will make transitions to the $L \sim 0$ final states (including the ground or Laughlin state), and the second kind of states will make transitions to the final states near the roton minimum, and higher energy states with the same $L$. From the properties of the magneto-exciton, we know that there are three regions of $d$ in which the optical spectra can be classified:

(i) In the small $d$ limit, $d \ll d_c$. There is a sharp peak which comes from transition $0 \rightarrow 0$ (for $P_0(\omega)$) or $1 \rightarrow 0$ (for $P_1(\omega)$). This peak has its origin as discussed above for the symmetric model ($d = 0$). For this strong peak, the transition is from the first kind of states to the final state which is the ground state.

(ii) Increasing the e-h distance $d$, the intensity of this peak will decrease, due to the slow decrease of transition matrix element as well as increase of energy of initial states relative to that of the second kind of initial states. For finite $d$, a lower energy peak develops and its intensity increase as $d$ increases. This peak corresponds to the transition from the second kind of initial states to the roton minimum final states. Near the crossover region $d \sim d_c$, the two peaks have comparable strength, and the energy splitting is around the gap energy $\Delta$ of the Laughlin incompressible state.

(iii) If $d$ increases further, the high energy peak will disappear, only the low energy peak exists. This peak is due to the transition from the lowest energy branch to the roton minimum states, so it is a "broad" peak. For $d > 2R$, there will be another lower energy peak [7,18] due to the transition to higher energy final states. Since the doublets (or multi-peaks) are from the same kind of initial states, the temperature dependence of different peaks are insignificant, in contrast to what was found in experiments (see below). Also because the stereographic meaning of $d > 2R$ is unclear, we do not show PL spectra for $d > 2R$. The detailed discussion of the PL spectra will be given elsewhere [14].

The PL spectra are shown in Fig.[2] for $P_0(\omega)$ and $P_1(\omega)$ respectively. The spectra in the figures were smoothed by hand using broadening of the delta functions to Lorentzian distribution with half-width $\gamma = 0.004e^2/\varepsilon l_0$. The difference between the structure of $P_1(\omega)$ spectra and $P_0(\omega)$ spectra are rather insignificant. The doublet structure appears in both $P_1(\omega)$ and $P_0(\omega)$.

In this calculation using spherical geometry, the doublets structure happens only in the transition region $d \sim 0.8R$. The reported experiments [3] are in plane geometry, however we suggest that this critical region in spherical geometry corresponds to a very broad region in the flat plane geometry, most the experiments in which the distance $d$ between electron and valence hole is comparable to the magnetic length $l_0$ belong to this transition region. so the doublet structures can be observed. We should mention that the doublets have different temperature behavior for $d < d_c$ and $d > d_c$. From this, we can know which region the experiments are belong to. For example, in the experiment [3] the lower energy peak increases with the increase of temperature, this corresponds to $d < d_c$ [8]. However, since the distance between the electron and hole might also depend on the temperature in some of the experiments, and the doublets are very sensitive to the distance $d$ between electron and hole, the temperature dependence of the distance $d$ can have important effects on the PL spectra.

Next we calculate the inelastic interband RS using the parameters relevant to Pinczuk *et al.*'s experiment [4]. As for Eq.(2), the following equation for the RS efficiency ($Q = 0$) can be derived using the Fermi Golden rule:

$$\mathcal{W}(\Omega) = M \sum_{n,i,\alpha,\beta} e^{-\beta E_i^e} \delta(\Omega - E_i^e - E_n^e) \left| \sum_m \langle \Psi_i^e | \hat{\mathcal{L}}_\alpha^\dagger | \Psi_m^{ex} \rangle \frac{1}{\omega_1 + E_i^e - E_m^{ex} - i\gamma} \langle \Psi_m^{ex} | \hat{\mathcal{L}}_\beta | \Psi_n^e \rangle \right|^2 \quad (3)$$

Here $\gamma$ is the finite hole-lifetime broadening, and $\alpha$ and $\beta$ are the index of the (circular) polarization. We also renormalize (shift) the photon energy by the "activation" energy of the valence hole as we did in the calculation of PL spectrum.

Our calculation was again done using spherical geometry with $N = 5$ electron in the initial states at $\nu = 1/3$, and $N+1$ electrons plus 1 valence hole in the intermediate states. The same calculation has been done for $N = 4$ and the RS spectrum is basically the same. Since in the RS, the initial states are incompressible and $\Delta \gg T$, we will assume $T = 0$. And we will only consider the $\hat{\mathcal{L}}_1$ process since the experiments are done in backscattering geometry. The calculated RS spectrum for the scattering process $\alpha \rightarrow \beta$ with ($\alpha \neq \beta$) is shown in Fig.[3]. We found that for this process, in which the circular polarization of the incident and scattered photon are opposite, the RS has a robust peak at an energy slightly larger than the gap value $\Delta$. For $N = 5$, the gap is $\Delta = 0.09E_c$, while the resonance peak is at $\Omega = 0.12E_c$. This is not surprising since the magneto-roton at the minimum energy has wave vector around $k_0 \sim 1/(\nu l_0)$, and the resonance is at $Q = 0$. It is believed that the gap-excitation in the $k = 0$ limit can be approximated by a two-roton bound state [11]. Using the result of Haldane and Rezayi [19], $\Omega = 0.12E_c$ is around the gap value of gap-excitation at $k = 0$. Another feature in Fig.[3] is that the resonance only occurs in a narrow frequency region of the incident photon frequency. This agree with the experiments [4]. For the process that the incident and out scattered photon has the same (circular) polarization ($\alpha = \beta$), there is only an *elastic* peak. The resonance RS at $\nu = 2/5$ has also been calculated and no robust resonance peak close to gap value $\Delta_{2/5}$ can be found [14].

This work is supported in part by FRAP-PSC-CUNY program and NSF-INT-9122114. JZ is supported by a Mina Rees Dissertation Fellowship of CUNY Graduate School.

*Note added* —After we submitted this paper, we received a preprint on the spectrum of the magneto-exciton by Apalkov and Rashba, who obtained similar results on the magneto-exciton spectrum. We also noticed that in Ref. [18], PL spectra $P_0(\omega)$ was calculated at large distance $d \sim 2.9R$ (see our discussion (iii) in text).

## REFERENCES

* Present address: T-11 Group, Theoretical Division, LANL, Los Alamos, NM 87545.

FIGURES

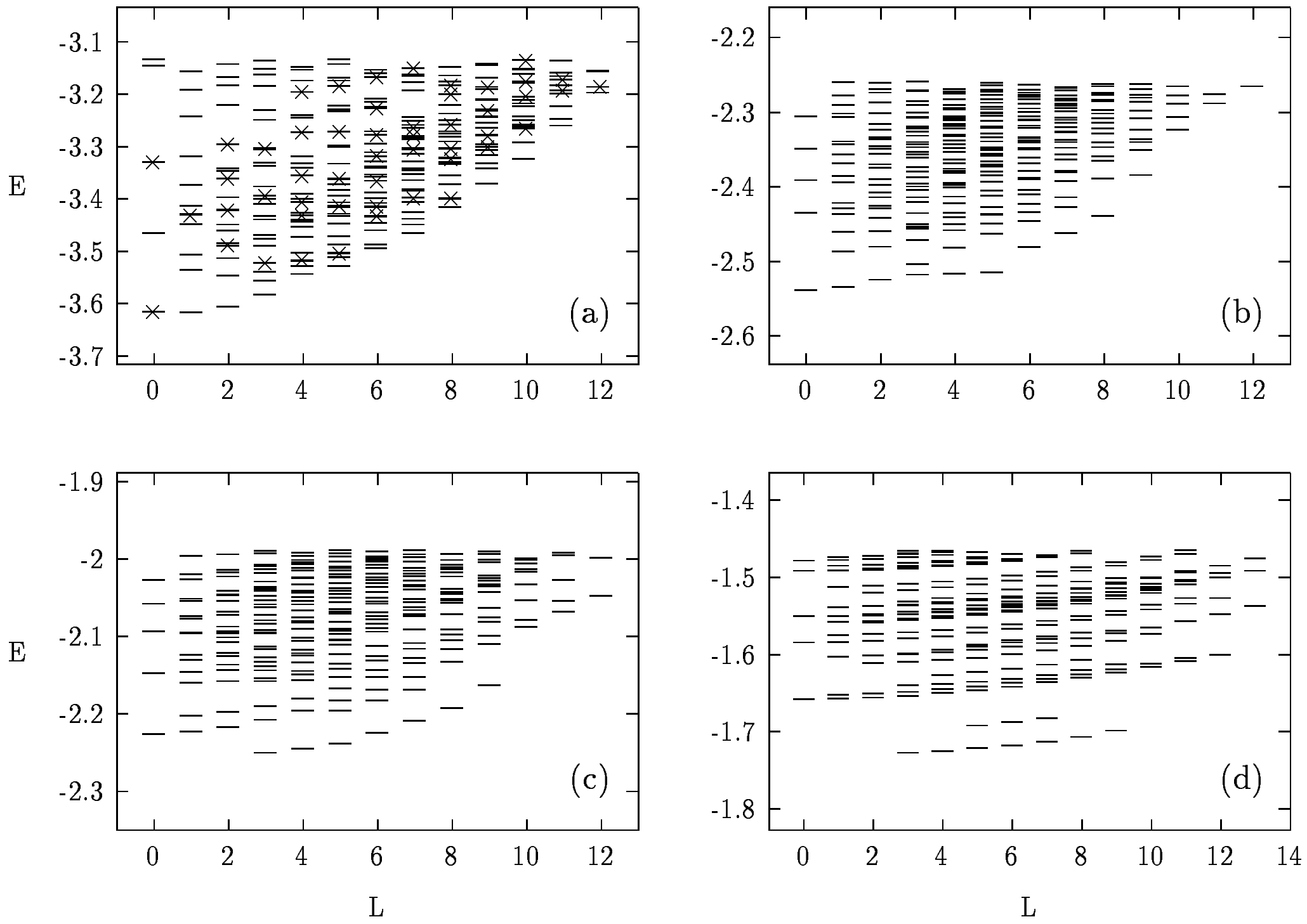

FIG. 1. Spectrum of the magneto-exciton for $2S = 12$ and 6e plus 1h. (a) $d = 0R$; (b) $d = 0.6R$; (c) $d = 0.9R$; (d) $d = 1.8R$. In (a), the crossed states correspond FQHE states at $\nu = 1/3$ (see text).

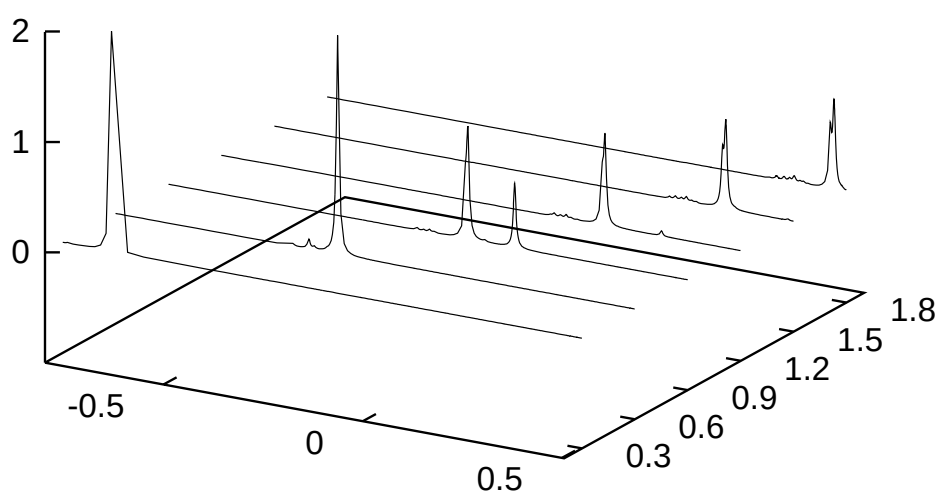

(a)

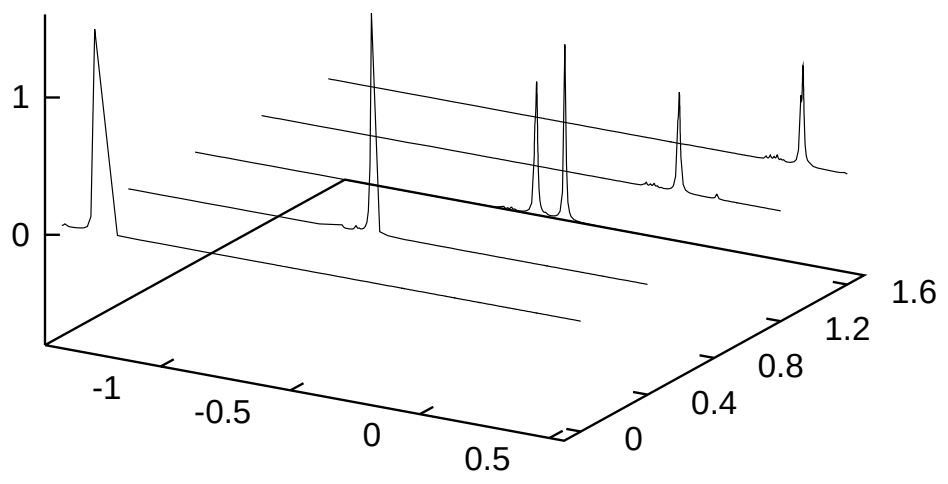

(b)

FIG. 2. PL spectra (a) $P_0(\omega)$ and (b) $P_1(\omega)$ as function of $d$ at $T = 0.005$. The peaks for (a) $d = 0.3R, 0.6R$; (b) $0.3R$ has been reduced by factor of 4. The $\omega$ unit of $x$-axis is $e^2/\varepsilon l_0$ and the distance $d$ unit of $y$-axis is $R$. Note the apparence of the doublet in the critical region $d \sim d_c$.

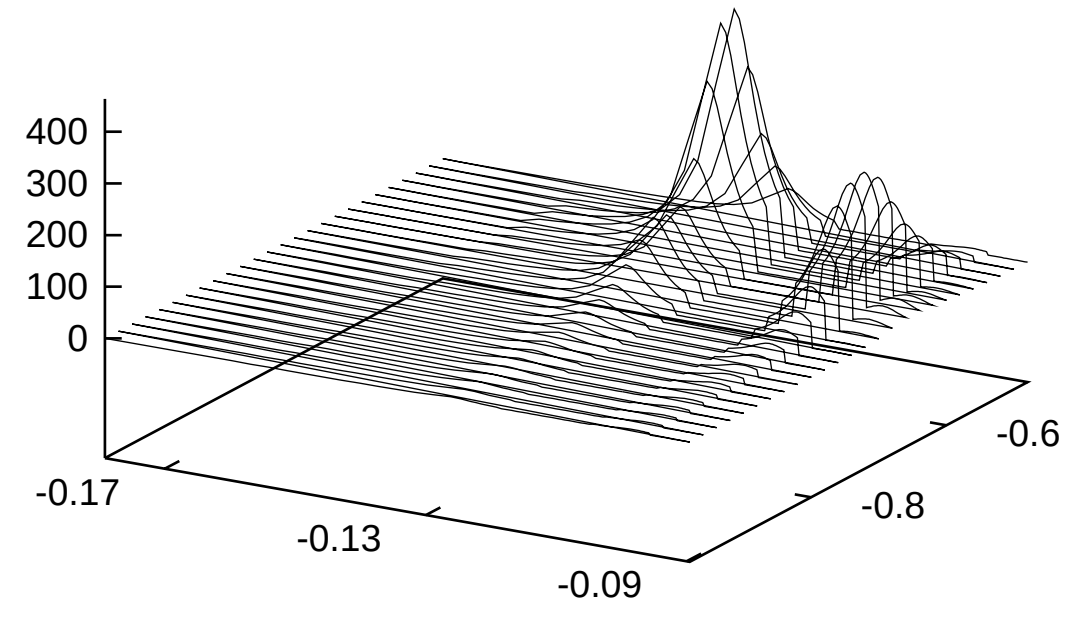

(a)

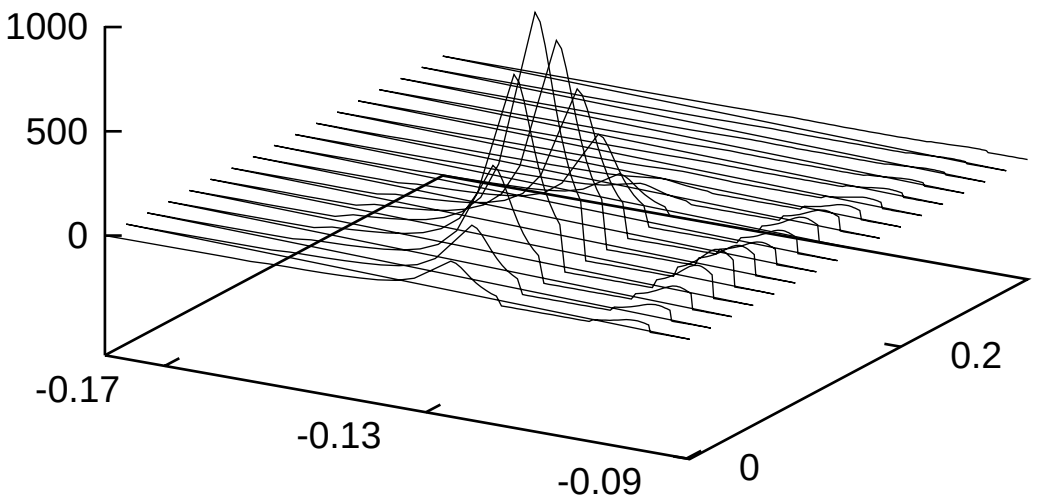

(b)

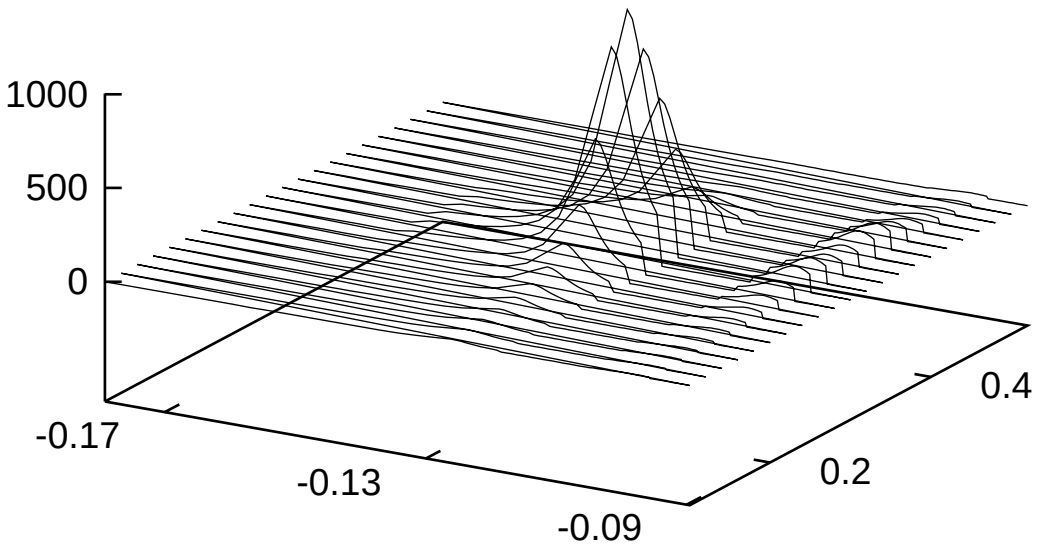

(c)

FIG. 3. RS spectrum $\mathcal{W}(\Omega, \omega_1)$ for (a) $d = 0.3R$; (b) $d = 0.9R$; (c) $d = 1.2R$. $\gamma = 0.03$ The $\delta$ function in Eq.(3) is broadened to Lorentzian function with halfwidth $\Gamma = 0.03$.